\documentclass{llncs}
\usepackage{latexsym}
\usepackage[latin1]{inputenc}

\usepackage{amsmath}

\usepackage{color}
\usepackage{graphicx} 
\usepackage{colortbl}
\usepackage{rotating}
\usepackage{array}

\usepackage{citesort}

\newcommand{\ciaopp}{\texttt{CiaoPP}}

\newcommand{\internalcomment}[1]{ }

\newcommand{\headtimeconstant}{K} 
\newcommand{\constantstep}{\headtimeconstant_{\modelstep}} 
\newcommand{\constantnumargs}{\headtimeconstant_{\modelnumargs}} 

\newcommand{\constantnumgroundinputunifs}{\headtimeconstant_{\modelnumgroundinputunifs}}
\newcommand{\constantnumgroundoutputunifs}{\headtimeconstant_{\modelnumgroundoutputunifs}}
\newcommand{\constantnumvarinputunifs}{\headtimeconstant_{\modelnumvarinputunifs}}
\newcommand{\constantnumvaroutputunifs}{\headtimeconstant_{\modelnumvaroutputunifs}}
\newcommand{\headconstant}{\headtimeconstant_{\costmodel_i}}
\newcommand{\headconstantvector}{\overline{\headtimeconstant}_{\vectorcostmodel}}

\newcommand{\modelstep}{step} 
\newcommand{\modelnumargs}{nargs} 
\newcommand{\modelnumgroundinputunifs}{giunif}
\newcommand{\modelnumgroundoutputunifs}{gounif}
\newcommand{\modelnumvarinputunifs}{viunif}
\newcommand{\modelnumvaroutputunifs}{vounif}
 
\newcommand{\timeanalysis}{T_{ca}}

\newcommand{\vectorcostmodel}{\Omega}

\newcommand{\costmodel}{\omega}

\newcommand{\headcost}{\tau}

\newcommand{\headtime}{time}

\newcommand{\indepmetric}{I}

\newcommand{\indepmetricvector}{\overline{\indepmetric(\vectorcostmodel)}}

\newcommand{\platformintel}{intel}
\newcommand{\platformppc}{ppc}

\newcommand{\forasap}[1]{ }

\newcommand{\forlongvers}[1]{ }

\newcolumntype{I}{!{\vrule width 0.75pt}}
\newlength\savedwidth
\newcommand\whline{\noalign{\global\savedwidth\arrayrulewidth
    \global\arrayrulewidth 0.75pt}%
  \hline
  \noalign{\global\arrayrulewidth\savedwidth}}

\newcommand{\wcline}[2]{\noalign{\global\savedwidth\arrayrulewidth
    \global\arrayrulewidth 0.75pt}%
  \cline{#1-#2}
  \noalign{\global\arrayrulewidth\savedwidth}}

\newcommand{\numvaluesperexample}{10}
\newcommand{\callspermeasure}{5}

\begin{document}

\title{Towards Execution Time Estimation \\ 
       for Logic Programs \\
       via Static Analysis and Profiling
\\}

\author{Edison Mera\inst{1} \and Pedro López-García\inst{1}  \and Germ\'{a}n
  Puebla\inst{1} \and \\ Manuel Carro\inst{1} \and Manuel Hermenegildo\inst{1,2} }

\institute{Technical University of Madrid\\
\email{edison@clip.dia.fi.upm.es}, 
\email{\{pedro.lopez,german,mcarro,herme\}@fi.upm.es}
\and  
University of  New Mexico,
\email{herme@unm.edu}
}

\vspace{-0.5\baselineskip}

\maketitle

\begin{abstract}
  \vspace{-\baselineskip} Effective static analyses have been proposed
  which infer bounds on the number of resolutions or
  reductions. These have the advantage of being independent from the
  platform on which the programs are executed and have
  been shown to be useful in a number of applications, such as granularity
  control in parallel execution. On the other hand, in 
  distributed computation scenarios where platforms with different
  capabilities come into play, it
  is necessary to express costs in metrics that include the
  characteristics of the platform. In particular, it is specially
  interesting to be able to infer upper and lower bounds on actual
  execution times.  With this objective in mind, we propose an approach
  which combines compile-time  analysis for cost bounds with a one-time
  profiling of the platform in order to determine the values of
  certain parameters for a given platform. These parameters calibrate a
  cost model which, from then on, is able to compute statically time
  bound functions for procedures and to predict with a significant
  degree of accuracy the execution times of such procedures in the
  given platform. The approach has been implemented and integrated in
  the \ciaopp\ system.

  \ \\
  \textbf{Keywords:} Execution Time Estimation, Cost Analysis,
  Profiling, Resource Awareness, Cost Models, Mobile Computing.

\end{abstract}

\section{Introduction}
\label{introduction}

Predicting statically the running time of programs has many
applications ranging from task scheduling in parallel execution to
proving the ability of a program to meet strict time constraints in
real-time systems.  A starting point in order to attack this problem
is to infer the computational complexity of such programs.
This is one of the reasons why the development of static analysis
techniques for inferring cost-related properties of programs has
received considerable attention.
However, in most cases such cost properties are expressed using
platform-independent metrics. For example,~\cite{granularity,caslog}
present a method for automatically inferring functions which capture
an upper bound on the number of resolution steps or reductions that a
procedure will execute as a function of the size of its input data.
In~\cite{granularity-jsc-short,pedro-phd} the method
of~\cite{granularity,pedro-phd} was fully automated in the context of
a practical compiler and in~\cite{low-bounds-ilps97,pedro-phd} a
similar approach was applied in order to also obtain lower bounds,
which are specially relevant in parallel execution.  Such
platform-independent cost information (bounds on number of reductions)
has been shown to be quite useful in various applications. This
includes, for example, scheduling parallel
tasks~\cite{pedro-phd,granularity-jsc-short,acc-res-ppdp05}.  In a
typical scenario, these tasks will be executed in a single parallel
machine, where all processors are typically identical. Therefore, the
deduced number of reductions can actually be used as a relative
measure in order to compare to a first degree of approximation the
amount of work under the tasks.

However, in distributed execution and other mobile/pervasive
computation scenarios, where different platforms come into play with
each platform having different computing power, it becomes necessary
to express costs in metrics that can be later instantiated to
different architectures so that actual running time can be compared
using the same units. This applies also to heterogeneous parallel
computing platforms.
With this objective in mind, we present a framework which combines
cost analysis with profiling techniques in order to infer functions
which yield bounds on platform-dependent \emph{execution times} of
procedures. 
Platform-independent cost functions are first inferred which are
parameterized by certain constants. These constants aim at capturing
the execution time of certain low-level operations on each platform.
For each execution platform, the value of such constants is determined
experimentally once and for all by running a set of 
synthetic benchmarks and measuring their running times with a
profiling toolkit that we have also developed.  Once these constants
are determined, they are fed into the model with the objective of
predicting with a certain accuracy execution times.  We have studied a
relatively large number of cost models, involving different sets of
constants in order to explore experimentally
which of the models produces the most precise results, i.e., which
parameters model and predict best the actual execution times of
procedures.  In doing this we have taken into account the trade-off
between simplicity of the cost models (which implies efficiency of the
cost analysis and also simpler profiling) and the precision of their
results. With this aim, we have started with a simple model and
explored several possible refinements.

In addition to cost analysis, the implementation of profilers in
declarative languages has also been considered by various authors,
with the aim of helping to discover why a part
of a program does not exhibit the expected performance.
Debray~\cite{Debray83} showed the basic considerations to have in mind
when profiling Prolog programs: handling backtracking and failure.
Ducass\'{e}~\cite{Ducasse99} designed and implemented  a trace
analyzer for Prolog which can be applied to profiling.
Sansom and Peyton Jones~\cite{sansom97formally} focused on profiling
of functional 
languages using a semantic approach and highlighted the difficulty in
profiling such kind of languages.  Jarvis and Morgan~\cite{morgan98} showed how
to profile lazy functional programs.
Brassel et al.~\cite{Brassel04LOPSTR} solved part of the difficulty in
profiling when considering special features in functional logic
programs, like sharing, laziness and non-determinism.  
We will use also profiling but, since our aim is to \emph{predict}
performance, profiling will in our case be aimed at calibrating the
values for some constants that appear in the cost functions, and which
will be instrumental to forecast execution times for a given platform
and cost model.
Therefore we will not use profiling with just some fixed input
arguments, but with a set of programs and input arguments which we
hope will be representative enough to derive meaningful
characteristics of an execution platform.

\section{Static Platform-Dependent Cost Analysis}
\label{sec:cost-analysis-approach}

In this Section we present the compile-time cost bounds analysis
component of our combined framework.
This analysis has been implemented and integrated in
\ciaopp~\cite{ciaopp-sas03-journal-scp-short} by extending 
previous implementations of reduction-counting cost analyses.  The
inferred (upper or lower) bounds on cost are expressed as functions on
the sizes of the input arguments and use several platform-dependent
parameters. Once these parameters are instantiated
with values for a given platform, such functions yield bounds on the
execution times required by the computation on such platform.  The
analyzer can use several metrics for computing the ``size'' of an
input, such as list-length, term-size, term-depth, integer-value, etc.
Types, modes, and size measures are first automatically inferred by
other analyzers which are part of \ciaopp\ and then used in the size
and cost analysis.

\subsection{Platform-Independent Static Cost Analysis}
\label{sec:previous-cost-analysis}

As mentioned before, our static cost analysis approach is based on
that developed 
in~\cite{granularity,caslog} (for estimation of upper bounds on
resolution steps) and further extended in~\cite{low-bounds-ilps97}
(for lower bounds). In these approaches the time complexity of a
clause can be bounded by the time complexity of head unification
together with the time complexity of each of its body literals. For
simplicity, the discussion that follows is focused on the estimation
of upper bounds. We refer the reader to~\cite{low-bounds-ilps97} for
details on lower bounds analysis.
Consider a clause ${\tt C}$ defined as ``${\tt H} :- {\tt L_1}, ...,
{\tt L_m}$''.  
Because of backtracking, the number of times a literal will be
executed depends on the number of solutions that the literals
preceding it can generate. Assume that $\overline{n}$ is a vector such
that each element corresponds to the size of an input argument to
clause ${\tt C}$ and that each $\overline{n}_i$, $i=1\ldots m$, is a
vector such that each element corresponds to the size of an input
argument to literal ${\tt L_{\it i}}$, $\headcost$ is the cost needed
to resolve the head ${\tt H}$ of the clause with the literal being
solved, and ${\tt Sols_{L_{\it j}}}$ is the number of solutions
literal ${\tt L_{\it j}}$ can generate. Then, an upper bound on the
cost of clause
${\tt C}$ (assuming all solutions are required), ${\tt
  Cost_{C}}(\overline{n})$, can be expressed as:
\begin{equation}
{\tt Cost_{C}}(\overline{n}) \leq \headcost + \sum\limits_{i=1}^{m}
(\prod\limits_{j \prec i} {\tt Sols_{L_{\it j}}}(\overline{n}_j)) 
{\tt Cost_{L_{\it i}}}(\overline{n}_i),
\label{eq:clause}
\end{equation}
Here we use $j \prec i$ to denote that ${\tt L_{\it j}}$ precedes
${\tt L_{\it i}}$ in the literal dependency graph for the clause.

Our current implementation also considers the cost of the terms
created for the literals in the body of predicates, which can affect
the cost expression significantly.  
To further simplify the discussion that follows, we restrict ourselves
to the simple case where each literal is determinate, i.e., produces
at most one solution.
In this case, equation (\ref{eq:clause}) simplifies to:
\begin{equation}
{\tt Cost_{C}}(\overline{n}) 
 \leq \headcost + \sum\limits_{i=1}^{m} {\tt Cost_{L_{\it i}}}(\overline{n}_i).
\label{eq:clause:det}
\end{equation}

\noindent
(However, it is important to note that our implementation is not
limited to deterministic programs: our system handles non determinism,
i.e., presence of several solutions for a given call, in the cost
analysis).

A difference equation is set up for each recursive clause, whose
solution (using as boundary conditions the cost of 
non-recursive clauses) is a function that yields the cost of a clause.
The cost of a predicate is then computed from the cost of its defining
clauses.  Since the number of solutions generated by a predicate that
will be demanded is generally not known in advance, a conservative
upper bound on the computational cost of a predicate can be obtained
by assuming that all solutions are needed, and that all clauses are
executed (thus the cost of the predicate is assumed to be the sum of
the costs of its defining clauses).
Taking mutual exclusion into account in order to obtain a more precise
estimate of the cost of a predicate is relatively easy: the
complexity for deterministic predicates can be approximated with the
maximum of the costs of mutually exclusive groups of clauses.

The analysis in~\cite{granularity,caslog} was primarily aimed at
estimating resolution steps.  However, the basic metric is open and
can be tailored to alternative scenarios: more sophisticated, accurate
measures can be used instead of the initially proposed ones (e.g.,
number of \emph{basic} unifications).
In the rest of this section we explore this open issue more deeply and
study how the original cost analysis can be extended in order to infer
cost functions using more refined and parametric cost models, which in
turn will 
allow achieving accurate execution time bound analysis.

\subsection{Proposed Platform-Dependent Cost Analysis Models}
\label{sec:proposed-cost-models}

Since the cost metric which we want to use in our approach is
execution time, we take $\headcost$ (in
expression~\ref{eq:clause:det}) to include the time needed to resolve
the head ${\tt H}$ of the clause with the literal being solved, the
cost associated with the resolution of the clause, and the cost coming
from setting up the body literals for execution.  %
In the following, we will refer to $\headcost$ as the {\em clause head
  cost function}, under the assumption that these other costs are also
taken into account. We will consider different values for $\headcost$,
each of them yielding a different cost model.  These cost models make
use of a vector of platform-dependent constants, together with a
vector of platform-independent metrics, each one corresponding to a
particular low-level operation related to program execution.  Examples
of such low-level operations considered by the cost models are
unifications where one of the terms being unified is a variable and
thus behave as an ``assignment'', or full unifications, i.e., when
both terms being unified are not variables, and thus unification
performs a ``test'' or produces new terms, etc.  Thus, we assume that
$\headcost$ is a function parameterized by the cost model, so that:

\begin{equation}
\headcost(\vectorcostmodel) = \headtime(\vectorcostmodel)
\label{eq:headcostfunc}
\end{equation}

\noindent 
where $\headtime(\vectorcostmodel)$ is a function that gives the time
needed to resolve the head ${\tt H}$ of the clause with the literal
being solved (plus some possible costs associated to the execution of
the clause such as, e.g., whether an activation record is allocated)
for the cost model named $\vectorcostmodel$. We study a family of cost
models such that $\headtime(\vectorcostmodel)$ is a function defined as
follows:

\begin{equation}
\headtime(\vectorcostmodel) = \headtime(\costmodel_1) + \cdots +
\headtime(\costmodel_v), \ v > 0 
\label{eq:vectorheadtimefunc}
\end{equation}

\noindent
where each $\headtime(\costmodel_i)$ provides that part of the execution time
which depends on the metric $\costmodel_i$. We assume that:

\begin{equation}
\headtime(\costmodel_i) = \headconstant \times \indepmetric(\costmodel_i)
\label{eq:headtimefunc}
\end{equation}

\noindent
where $\headconstant$ is a platform-dependent constant, and
$\indepmetric(\costmodel_i)$ is a platform-independent cost function.

Since $\headtime(\vectorcostmodel)$ is a linear combination of
platform-independent cost functions, we can write
equation~(\ref{eq:vectorheadtimefunc}) as:

\begin{equation}
\headtime(\vectorcostmodel) = \headconstantvector \bullet \indepmetricvector
\label{eq:headtimefunc-vector}
\end{equation}
 
\noindent
where $\headconstantvector$ is a vector of platform-dependent
constants, $\indepmetricvector$ is a vector of platform-independent
cost functions, and $\bullet$ is the dot product.

Accordingly, we generalize the definition of
equation~(\ref{eq:clause:det}) introducing the clause head cost
function $\headcost$ as a parameter:

\begin{equation}
{\tt Cost_{C}}(\headcost,\overline{n}) \leq \headcost +
\sum\limits_{i=1}^{m} 
{\tt Cost_{L_{\it i}}}(\overline{n}_i).
\label{eq:clause:param}
\end{equation}

A particular definition of $\indepmetricvector$ yields a cost model.
We have tried several cost models, by using different vectors
$\indepmetricvector$ constructed by choosing some (or all) of the
following $\indepmetric(\costmodel_i)$ cost functions (for example,
the cost model that uses \emph{all} such functions is
$\indepmetricvector = (\indepmetric(\modelstep),
\indepmetric(\modelnumvarinputunifs),
\indepmetric(\modelnumvaroutputunifs),
\indepmetric(\modelnumgroundinputunifs),
\indepmetric(\modelnumgroundoutputunifs))$).  In the following an
\emph{input argument} is one for which the term being passed by the
calling literal is known to be non-var at the time of head
unification. An \emph{output argument} is one for which the term being
passed by the calling literal is known to be a variable at the time
of head unification. Whether unifications are input or output can be
inferred using well-known techniques for mode analyses (in our case,
those provided by \ciaopp).  

\begin{itemize}
  
\item
  $\indepmetric(\modelstep) = 1$. \\
  Here we assume that there is a constant component of the execution
  time when a clause is resolved (a clause neck ``\texttt{:-}'' is
  crossed). I.e., following equation~(\ref{eq:headtimefunc}), we are
  assuming for this component that:

\[
\headtime(\modelstep) = \constantstep
\]

\item $\indepmetric(\modelnumvaroutputunifs) =$ {\em the number of
    variables in the clause head  which correspond to ``output'' argument positions.} \\
  Here we assume that there is a component of the execution time that
  is directly proportional to the number of cases where we know that
  both terms being unified are variables and thus unification really
  implies a simple assignment with a (presumably small) constant cost:

\[
\headtime(\modelnumvaroutputunifs) = \constantnumvaroutputunifs \times  \indepmetric(\modelnumvaroutputunifs) 
\]

\item $\indepmetric(\modelnumvarinputunifs) =$ {\em the number of
    variables in the clause head  which correspond to ``input'' argument positions}. \\
  Here we assume that there is a component of the execution time that
  is directly proportional to the number of cases where we know that
  the incoming term is non-var and the argument position in the clause
  is a variable. In this case the head unification for that argument
  is also an assignment with a small, constant cost, and there is also
  a cost associated with creating the input argument at the calling
  point, which for simplicity we will also consider constant. Given
  these assumptions:

\[
\headtime(\modelnumvarinputunifs) = \constantnumvarinputunifs \times  \indepmetric(\modelnumvarinputunifs) 
\]

\item $\indepmetric(\modelnumgroundoutputunifs) =$ {\em The number of
    function symbols, constants, and variables in the clause head
    which appear in output
    arguments.} \\
  We are assuming that there is a component of the execution time that
  is directly proportional to the size of the terms that have to be
  written into variables passed in by the calling literal, and which
  is proportional to the number of function symbols,
  constants, and variables which appear in output arguments in the
  clause head:

\[
\headtime(\modelnumgroundoutputunifs) = \constantnumgroundoutputunifs \times  \indepmetric(\modelnumgroundoutputunifs) 
\]

\item $\indepmetric(\modelnumgroundinputunifs) =$ {\em The number of
    function symbols, variables, and constants in the clause head
    which appear in input arguments.}\\
  Here we are assuming that there is a component of the execution time
  that is directly proportional to the number of ``input''
  unifications, i.e., when both terms being unified are not variables,
  and thus unification performs a ``test,'' and which is actually
  proportional to the number of function symbols, variables, and
  constants in the clause head which appear in input arguments (this
  is obviously an approximation):

\[
\headtime(\modelnumgroundinputunifs) = \constantnumgroundinputunifs \times  \indepmetric(\modelnumgroundinputunifs) 
\]

\item $\indepmetric(\modelnumargs) = arity(H)$. \\
  Here we are assuming that there is a component of the execution
  time that depends on the number of arguments in the clause head:

\begin{equation}
\headtime(\modelnumargs) = \constantnumargs \times  arity(H) 
\label{eq:numargsheadtime}
\end{equation}

This component is obviously redundant with respect to the previous
ones, but we have included it as a statistical control: the
experiments should show (and do show) that it is irrelevant when the
others are used. 
\ \\

\end{itemize}

Clearly, other components can be included (such as whether activation
records are created or not) but our objective is to see how far we can
go with the components outlined above.

We adopt the same approach as~\cite{caslog,low-bounds-ilps97} for
computing bounds on cost of predicates from the computed values for
the cost of the clauses defining it. However, we introduce the clause
head cost function $\headcost$ as a parameter of these cost
functions.

Let ${\tt Cost_p}(\headcost, \overline{n})$ be a function which gives
the cost of the computation of a call to predicate {\tt p} for an
input of size $\overline{n}$ (recall that the cost units depend on
the definition of $\headcost$).
Given a predicate {\tt p}, and a clause head cost function
$\headtime(\vectorcostmodel)$ of the form defined in
equation~(\ref{eq:headtimefunc-vector}), we have that:

\begin{equation}
{\tt Cost_p}(\headtime(\vectorcostmodel), \overline{n}) = \headconstantvector
\bullet \overline{{\tt Cost_p}}(\indepmetricvector, \overline{n})
\label{eq:costrelation-vector}
\end{equation}

\noindent
where $\headconstantvector$, $\indepmetricvector$ and $\overline{{\tt
    Cost_p}}(\indepmetricvector, \overline{n})$ are vectors of the
form:

$\headconstantvector = (\headtimeconstant_{\costmodel_1}, \ldots, \headtimeconstant_{\costmodel_v})$,

$\indepmetricvector = (\indepmetric(\costmodel_i), \ldots,
\indepmetric(\costmodel_v))$, and

$\overline{{\tt Cost_p}}(\indepmetricvector, \overline{n}) = ({\tt
  Cost_p}(\indepmetric(\costmodel_1), \overline{n}), \ldots, {\tt
  Cost_p}(\indepmetric(\costmodel_v), \overline{n}))$

\medskip

Equation~(\ref{eq:costrelation-vector}) 
gives the basis for computing values
for constants $\headconstant$ via profiling (as explained in Section
~\ref{sec:calibration-procedure}). Also, it provides a way to obtain
the cost of a procedure expressed in a platform-dependent cost metric
from another cost expressed in a platform-independent cost metric.

\section{Refining the Cost Model: Dealing with Builtins}
\label{sec:dealing-with-builtins}

In this section we present our approach to the cost analysis of
programs which call builtins, or more generally, predicates whose code
is not available to the analyzer (external predicates). We will refer
to all of them as builtins for brevity.
We assume that there is a cost function (expressed via trust
assertions~\cite{ciaopp-sas03-journal-scp-short}) for builtin
predicates. In some cases, this cost function for each builtin
predicate is approximated by a constant value, and in others, it is
approximated by a function that depends on properties of the (input)
arguments of the predicate. In particular, the cost of arithmetic
builtin predicates (such as {\tt =:=/2}, {\tt =\verb.\.=/2}, or {\tt
  >/2}) is approximated by a function that depends on the number and
type of arithmetic operands appearing in the arithmetic expressions
that can be passed to such predicates as arguments.

Note that this is an important improvement over the cost analysis
proposed in~\cite{caslog} (which infers {\em number of resolution
  steps}), since one of the assumptions made in such analysis is that
calls to certain builtin predicates are not counted as a resolution
step, and are thus completely ignored by cost analysis. This
assumption is not realistic if we want to estimate execution times,
since the cost of executing such builtins has to be taken into account.

Going into more detail, we assume that each builtin contributes with a
new component to the 
execution time as expressed in Equation~(\ref{eq:vectorheadtimefunc}),
that is, our cost model will have a new component
$\headtime(\costmodel_i)$ for each builtin predicate and arithmetic
operator. Let $\odot/n$ be an arithmetic operator. The execution time
due to the total number of times that such operator is evaluated is
given by:

\[
\headtime(\odot/n) = \headtimeconstant_{\odot/n} \times \indepmetric(\odot/n)
\]
 
\noindent
where $\headtimeconstant_{\odot/n}$ is a platform-dependent constant,
and $\indepmetric(\odot/n)$ is a platform-independent cost function.
$\headtimeconstant_{\odot/n}$ approximates the cost (in units of time)
of evaluating the arithmetic operator $\odot/n$.
$\indepmetric(\odot/n)$ could be the number of times that the
arithmetic operator is evaluated. Alternatively, it can be a cost
function defined as:

\[
 I(\odot/n) = \sum\limits_{a \in S}{\tt EvCost}(\odot/n, a)
\]

\noindent
and where $S$ is the set of arithmetic expressions appearing in the clause
body which will be evaluated; and ${\tt EvCost}(\odot/n, a)$
represents the cost corresponding to the operator ${\odot/n}$ in the
evaluation of the arithmetic term $a$, i.e.:

\[
  {\tt EvCost}(\odot/n,A) = 
  \begin{cases}
    0 & \text{if $A$ is a constant}\\
      & \text{~~~~or a variable} \\
    1 + \sum\limits_{i=1}^{n}{{\tt EvCost}(\odot/n, A_{i})}
    & \text{if $A = \odot(A_1, ..., A_n)$} \\
    \sum\limits_{i=1}^{m}{{\tt EvCost}(\odot/n, A_{i})}
    & \text{if $A \neq \odot(A_1, ..., A_n)$} \\
    & \wedge \text{ $A = \hat{\odot}(A_1, ..., A_m)$}\\
    & \text{for some operator ${\hat{\odot}/m}$}
  \end{cases}
\]

\noindent

For simplicity, we assume that the cost of evaluating the arithmetic
term $t$ to which a variable appearing in $A$ will be bound at
execution time is zero (i.e., we ignore the cost of evaluating $t$).
This is a good approximation if in most cases $t$ is a number and
thus no evaluation is needed for it.  However, a more refined cost
model could assume that this cost is a function on the size of $t$.

Note that this model ignores the possible optimizations that the
compiler might perform.  We can take into account those performed by
source-to-source transformation by placing our analyses in the last
stage of the front-end, but at some point the language the compiler
works with would be different enough as to require different
considerations in the cost model.

However, exp\-er\-i\-men\-tal results show
that our simplified cost model gives a good approximation of the
execution times for arithmetic builtin predicates.
With these assumptions, equation~(\ref{eq:costrelation-vector})
(in Section~\ref{sec:proposed-cost-models}) also holds for programs
that perform calls to builtin predicates, say, for example, a builtin
$b/n$, by introducing $b/n$ and $\odot/n$ as new cost components of
$\Omega$.

A similar approach can be used for other (non-arithmetic) builtins
$b/n$ using the formula:

\[
\headtime(b/n) = \headtimeconstant_{b/n} \times \indepmetric(b/n)
\]

\section{Calibrating Constants via Profiling}
\label{sec:calibration-procedure}

In order to compute values for the platform-dependent constants which
appear in the different cost models
proposed in Section~\ref{sec:proposed-cost-models}, our calibration
schema takes advantage of the relationship between the platform-dependent
and -independent cost metrics expressed in
Equation~(\ref{eq:costrelation-vector}).
In this sense, the calibration of the constants appearing in
$\headconstantvector$ is performed by solving systems of linear
equations (in which such constants are treated as variables).

\noindent
Based on this expression, the calibration procedure consists of:

\begin{enumerate}
  
\item Using a selected set of calibration programs which
  aim at isolating specific aspects that affect execution time of
  programs in general.  For these calibration programs it holds that
  ${\tt Cost_p}(\indepmetric(\costmodel_i), \overline{n})$ is known
  for all $1 \leq i \leq v$. This can be done by using any of the
  following methods:

  \begin{itemize}
  
  \item The analyzers integrated in the \ciaopp\ system infer the
    exact cost function, i.e., ${\tt
      Cost_{p}}^{l}(\indepmetric(\costmodel_i), \overline{n}) = {\tt
      Cost_{p}}^{u}(\indepmetric(\costmodel_i), \overline{n}) = {\tt
      Cost_p}(\indepmetric(\costmodel_i), \overline{n})$ ,
    
  \item ${\tt Cost_p}(\indepmetric(\costmodel_i), \overline{n})$ is 
    computed by a profiler tool, or

  \item ${\tt Cost_p}(\indepmetric(\costmodel_i), \overline{n})$ is
    supplied by the user together with the code of program ${\tt p}$
    (i.e., the cost function is not the result from any automatic analysis
    but rather ${\tt p}$ is well known and its cost function can be
    supplied in a trust assertion).

\end{itemize}
  
\item For each benchmark $p$ in this set, automatically generating a
  significant amount $m$ of input data for it. This can be achieved by
  associating with each calibration program a data generation rule.
  
\item For each generated input data $d_j$, computing a pair $(\overline{C}_{p_j}, T_{p_j})$,
  $1 \leq j \leq m$, where:

   \begin{itemize}
     
   \item $T_{p_j}$ is the $j$-th observed execution time of program $p$ with
     this generated input data. 
     
   \item $\overline{C}_{p_j} = \overline{{\tt
         Cost_p}}(\indepmetricvector, \overline{n_j})$, 
         where $\overline{n_j}$ is the size of the
         $j$-th input data $d_j$.

   \end{itemize}
   
 \item Using the set of pairs $(\overline{C}_{p_j}, T_{p_j})$ for
     setting up the equation:

\begin{equation}
\overline{C}_{p_j} \bullet \headconstantvector = T_{p_j}
\label{eq:vectorcalibration}
\end{equation}

where $\headconstantvector$ is considered a vector of variables.

\item Setting up the (overdetermined) system of equations composed by
  putting together all the equations~(\ref{eq:vectorcalibration})
  corresponding to all the calibration programs.
 
\item Solving the above system of equations using the least square
  method (see, e.g.,~\cite{mendenhall95}). A solution to this system
  gives values to the vector $\headconstantvector$ and hence, to the
  constants $\headconstant$ which are the elements composing it.

\item Calculating the constants for builtins and arithmetic
  operators by performing repeated tests in which only the builtin being
  tested is called, accumulating the time, and dividing the accumulated
  time by the number of times the repeated test has been performed.

\end{enumerate}

\section{Assessment of the Calibration of Constants}
\label{sec:assessment-initial-cost-model}



We have assessed both the constant calibration process and the
prediction of execution times using the previously proposed cost
models in two different platforms:

\begin{itemize}
  
\item ``\platformintel'' platform: Dell Optiplex, Pentium 4 (Hyper
  threading), 2GHz, 512MB RAM memory, Fedora Core 4 operating System
  with Kernel 2.6.
  
\item ``\platformppc''  platform: Apple iMac, PowerPC G4 (1.1)
  1.5GHz, 1GB RAM memory, with Mac OS X 10.4.5 Tiger.

\end{itemize}

\forlongvers{
In Intel we are using the RDTSC instruction previously explained, and
in PPC the wall-time.  In both cases no other process where running
except the programs being bench-marked.
}



\begin{table}[t]
  \begin{center}
    \begin{tabular}{Il|l|cI} \whline
      Program \\ \whline
      Environments creation \\
      Predicates with no arguments \\
      Traverse a list without last call optimization \\
      Traverse a list with last call optimization \\
      Program for which $\indepmetric(\modelnumvarinputunifs)$ is
      known \\
      Program for which $\indepmetric(\modelnumvaroutputunifs)$ is
      known \\
      Program (unifying deep terms) for which $\indepmetric(\modelnumgroundinputunifs)$ is
      known \\
      Program (unifying flat terms) for which $\indepmetric(\modelnumgroundinputunifs)$ is
      known \\
      Program for which $\indepmetric(\modelnumgroundoutputunifs)$ is
      known \\
      Predicate with many arguments \\
      \whline
    \end{tabular}
  \end{center}
  \caption{Description of calibration programs used in the estimation of constants.}
  \label{k-calibration-progs}
  \vspace{-8mm}
\end{table}

In section~\ref{sec:calibration-procedure} we presented
equation~\ref{eq:vectorcalibration}, and we mentioned that it can be
solved using the least squares method.  We used the householder
algorithm, which consists in decomposing the matrix $C =
\{\overline{C}_{p_j}\}$, which has $m$ rows and $n$ columns into the
product of two matrices $Q$ and $U$ (denoted $ \bullet $ or without
any symbol) such that $C = Q \bullet U $, where $Q$ is an orthonormal
matrix (i.e., ${Q}^{T} \bullet Q = I$, the $m \times m$ identity
matrix) and $U$ an 
upper triangular $m \times n$ matrix.
Then, multiplying both sides of the
equation~\ref{eq:vectorcalibration} by $Q^{T}$ and simplifying we can
get:

\[
U \bullet K = Q^{T} \bullet T = B
\]

\noindent
where, for clarity, we denote $K = \headconstantvector$,
$T = T_{p_j}$ and $Q^{T} \bullet T = B$. We can take advantage of the
structure of $U$ and define $V$ as the first $n$ rows of $U$, $n$
being the number of columns of $C$ and $b$ the first $n$ rows of $B$,
then $K$ can be estimated solving the following upper triangular
system, where $\hat{K}$ stands for the estimate for $K$:

\[
V \bullet \hat{K} = Q^{T} \bullet T = b
\]

Since this method is being used to find an approximate solution, we
define the residual of the system as the value

\[
R = T - C \hat{K}
\]

\noindent
Let
\[
  RSS = R \bullet R
\]

\noindent
be the residual square sum, and let

\[
  MRSS = \frac{RSS}{m - n} 
\]

\noindent
be the mean of residual square sum, where $m$ and $n$ are the number
of rows and columns of the matrix $C$ respectively, and finally let

\[
  S = \sqrt{MRSS}
\]

\begin{table}[htbp]
  \begin{center}
    \begin{tabular}{Il|lIr|lI} \whline
      Plat. & Model & $S$ ($\mu s$) & \multicolumn{1}{cI}{${\headconstantvector}$} \\ \whline
      {\platformintel} & {\modelstep} {\modelnumargs} {\modelnumgroundinputunifs} {\modelnumgroundoutputunifs} {\modelnumvarinputunifs} {\modelnumvaroutputunifs}  & 6.2475 & (21.27, 9.96, 10.30, 8.23, 6.46, 5.69) \\
      & {\modelstep} {\modelnumgroundinputunifs} {\modelnumgroundoutputunifs} {\modelnumvarinputunifs} {\modelnumvaroutputunifs}  & 9.3715 & (26.56, 10.81, 8.60, 6.17, 6.39) \\
      & {\modelstep} {\modelnumgroundinputunifs} {\modelnumgroundoutputunifs} {\modelnumvaroutputunifs}  & 13.7277 & (27.95, 11.09, 8.77, 7.40) \\
      & {\modelstep}  & 68.3088 & 108.90 \\
      \hline
      {\platformppc} & {\modelstep} {\modelnumargs} {\modelnumgroundinputunifs} {\modelnumgroundoutputunifs} {\modelnumvarinputunifs} {\modelnumvaroutputunifs}  & 4.7167 & (41.06, 5.21, 16.85, 15.14, 9.58, 9.92) \\
      & {\modelstep} {\modelnumgroundinputunifs} {\modelnumgroundoutputunifs} {\modelnumvarinputunifs} {\modelnumvaroutputunifs}  & 5.9676 & (43.83, 17.12, 15.33, 9.43, 10.29) \\
      & {\modelstep} {\modelnumgroundinputunifs} {\modelnumgroundoutputunifs} {\modelnumvaroutputunifs}  & 16.4511 & (45.95, 17.55, 15.59, 11.82) \\
      & {\modelstep}  & 116.0289 & 183.83 \\
      \whline
    \end{tabular}
  \end{center}
  \caption{Global values for vector constants in several cost models
    (in nanoseconds), sorted by $S$, the standard error of the model.}
  \label{average-k-only-unif}
  \vspace{-3mm}
\end{table}

\noindent
be the estimation of the model standard error, $S$.
In order to experimentally evaluate which models better approximate the
observed time in practice, we have compared the values of $MRSS$ (or
$S$) for several proposed models. 
Table~\ref{average-k-only-unif} shows the estimated values for the
vector $K$ using the calibration programs in
Table~\ref{k-calibration-progs}, as well as the standard error of the
model, sorted from the best to the worst model.  For example, the
first row in the table shows the model that has as components
{\modelstep}, {\modelnumargs}, {\modelnumgroundinputunifs},
{\modelnumgroundoutputunifs}, {\modelnumvarinputunifs},
{\modelnumvaroutputunifs} for the \platformintel\ platform. It has a
standard error of 6.2475 $\mu s$ and the values for each of the constants
are 21.27, 9.96, 10.30, 8.23, 6.46, and 5.69 nanoseconds, respectively.

Note that the estimation of $K$ is done just once per platform.  In
the case of the \platformintel\ platform it took 15.62 seconds and in
\platformppc\ 17.84 seconds, repeating the experiment 250 times for
each program.

\section{Assessment of the Prediction of Execution Times}
\label{sec:asses:prediction}





\begin{table}[htbp]
  \begin{small}
  \begin{center}
\begin{tabular}{IlIlIc rIc rI} \whline
  Prog. & Model & \multicolumn{2}{cI}{{\platformintel}} & \multicolumn{2}{cI}{{\platformppc}} \\
 \cline{3-6}
  &  & \multicolumn{2}{cI}{Estimate}  & \multicolumn{2}{cI}{Estimate}  \\
 \cline{3-6}
  &  & ($\mu s$) & (\%)  & ($\mu s$) & (\%)  \\
 \whline
   \whline
  evpol & {\modelstep} {\modelnumargs} {\modelnumgroundinputunifs} {\modelnumgroundoutputunifs} {\modelnumvarinputunifs} {\modelnumvaroutputunifs}  & $89.72$ & ($44$) & $77.4$ & ($23$) \\
  & {\modelstep} {\modelnumgroundinputunifs} {\modelnumgroundoutputunifs} {\modelnumvarinputunifs} {\modelnumvaroutputunifs}  & $85.06$ & ($38$) & $74.96$ & ($26$) \\
  & {\modelstep} {\modelnumgroundinputunifs} {\modelnumgroundoutputunifs} {\modelnumvaroutputunifs}  & $82$ & ($35$) & $70.28$ & ($33$) \\
  & {\modelstep}  & $90.12$ & ($45$) & $85.07$ & ($13$) \\
  \cline{2-6}
  & Observed & $58.43$ & & $97.08$ & \\
  \cline{2-6}
  & Analysis time ${\timeanalysis}$ (s) & $2.002$ &  & $4.461$ &  \\
  \wcline{1}{6}
  \whline
  hanoi & {\modelstep} {\modelnumargs} {\modelnumgroundinputunifs} {\modelnumgroundoutputunifs} {\modelnumvarinputunifs} {\modelnumvaroutputunifs}  & $319$ & ($31$) & $398.5$ & ($4$) \\
  & {\modelstep} {\modelnumgroundinputunifs} {\modelnumgroundoutputunifs} {\modelnumvarinputunifs} {\modelnumvaroutputunifs}  & $243.3$ & ($3$) & $358.8$ & ($7$) \\
  & {\modelstep} {\modelnumgroundinputunifs} {\modelnumgroundoutputunifs} {\modelnumvaroutputunifs}  & $205.6$ & ($14$) & $301.3$ & ($25$) \\
  & {\modelstep}  & $340.7$ & ($38$) & $538.6$ & ($34$) \\
  \cline{2-6}
  & Observed & $235.3$ & & $384.2$ & \\
  \cline{2-6}
  & Analysis time ${\timeanalysis}$ (s) & $2.145$ &  & $4.903$ &  \\
  \wcline{1}{6}
  \whline
   \whline
  nrev & {\modelstep} {\modelnumargs} {\modelnumgroundinputunifs} {\modelnumgroundoutputunifs} {\modelnumvarinputunifs} {\modelnumvaroutputunifs}  & $131.3$ & ($68$) & $179.4$ & ($26$) \\
  & {\modelstep} {\modelnumgroundinputunifs} {\modelnumgroundoutputunifs} {\modelnumvarinputunifs} {\modelnumvaroutputunifs}  & $101.1$ & ($39$) & $163.6$ & ($16$) \\
  & {\modelstep} {\modelnumgroundinputunifs} {\modelnumgroundoutputunifs} {\modelnumvaroutputunifs}  & $82.51$ & ($18$) & $135.2$ & ($3$) \\
  & {\modelstep}  & $144.4$ & ($80$) & $243.8$ & ($59$) \\
  \cline{2-6}
  & Observed & $69.25$ & & $139.2$ & \\
  \cline{2-6}
  & Analysis time ${\timeanalysis}$ (s) & $2.022$ &  & $4.691$ &  \\
  \wcline{1}{6}
  \whline
  palind & {\modelstep} {\modelnumargs} {\modelnumgroundinputunifs} {\modelnumgroundoutputunifs} {\modelnumvarinputunifs} {\modelnumvaroutputunifs}  & $131.8$ & ($18$) & $179.8$ & ($5$) \\
  & {\modelstep} {\modelnumgroundinputunifs} {\modelnumgroundoutputunifs} {\modelnumvarinputunifs} {\modelnumvaroutputunifs}  & $101$ & ($9$) & $163.7$ & ($5$) \\
  & {\modelstep} {\modelnumgroundinputunifs} {\modelnumgroundoutputunifs} {\modelnumvaroutputunifs}  & $86.91$ & ($24$) & $142.1$ & ($19$) \\
  & {\modelstep}  & $167.2$ & ($43$) & $282.2$ & ($52$) \\
  \cline{2-6}
  & Observed & $110$ & & $171.6$ & \\
  \cline{2-6}
  & Analysis time ${\timeanalysis}$ (s) & $2$ &  & $4.7$ &  \\
  \wcline{1}{6}
  \whline
  powset & {\modelstep} {\modelnumargs} {\modelnumgroundinputunifs} {\modelnumgroundoutputunifs} {\modelnumvarinputunifs} {\modelnumvaroutputunifs}  & $537.5$ & ($59$) & $727.9$ & ($17$) \\
  & {\modelstep} {\modelnumgroundinputunifs} {\modelnumgroundoutputunifs} {\modelnumvarinputunifs} {\modelnumvaroutputunifs}  & $404.5$ & ($28$) & $658.3$ & ($7$) \\
  & {\modelstep} {\modelnumgroundinputunifs} {\modelnumgroundoutputunifs} {\modelnumvaroutputunifs}  & $323.8$ & ($5$) & $534.9$ & ($14$) \\
  & {\modelstep}  & $448.7$ & ($38$) & $757.4$ & ($21$) \\
  \cline{2-6}
  & Observed & $308.2$ & & $615$ & \\
  \cline{2-6}
  & Analysis time ${\timeanalysis}$ (s) & $2.07$ &  & $4.636$ &  \\
  \wcline{1}{6}
  \whline
  append & {\modelstep} {\modelnumargs} {\modelnumgroundinputunifs} {\modelnumgroundoutputunifs} {\modelnumvarinputunifs} {\modelnumvaroutputunifs}  & $50.29$ & ($75$) & $68.72$ & ($24$) \\
  & {\modelstep} {\modelnumgroundinputunifs} {\modelnumgroundoutputunifs} {\modelnumvarinputunifs} {\modelnumvaroutputunifs}  & $38.69$ & ($44$) & $62.65$ & ($15$) \\
  & {\modelstep} {\modelnumgroundinputunifs} {\modelnumgroundoutputunifs} {\modelnumvaroutputunifs}  & $31.36$ & ($22$) & $51.45$ & ($5$) \\
  & {\modelstep}  & $54.56$ & ($85$) & $92.1$ & ($56$) \\
  \cline{2-6}
  & Observed & $25.16$ & & $53.92$ & \\
  \cline{2-6}
  & Analysis time ${\timeanalysis}$ (s) & $1.932$ &  & $4.441$ &  \\
  \wcline{1}{6}
  \whline
\end{tabular}
  \end{center}
  \end{small}
\caption{Evaluation of execution time predictions.}  \label{predict-mean-consider_builtins}
\end{table}

\begin{table}[htbp]
  \begin{center}
\begin{tabular}{Il|lIrI} \whline
Platform & Model & Error (\%) \\ \whline
{\platformintel}
 & {\modelstep} {\modelnumargs} {\modelnumgroundinputunifs} {\modelnumgroundoutputunifs} {\modelnumvarinputunifs} {\modelnumvaroutputunifs}  & $53.17$ \\
 & {\modelstep} {\modelnumgroundinputunifs} {\modelnumgroundoutputunifs} {\modelnumvarinputunifs} {\modelnumvaroutputunifs}  & $31.06$ \\
 & {\modelstep} {\modelnumgroundinputunifs} {\modelnumgroundoutputunifs} {\modelnumvaroutputunifs}  & $21.48$ \\
 & {\modelstep}  & $58.45$ \\
\hline
{\platformppc}
 & {\modelstep} {\modelnumargs} {\modelnumgroundinputunifs} {\modelnumgroundoutputunifs} {\modelnumvarinputunifs} {\modelnumvaroutputunifs}  & $18.72$ \\
 & {\modelstep} {\modelnumgroundinputunifs} {\modelnumgroundoutputunifs} {\modelnumvarinputunifs} {\modelnumvaroutputunifs}  & $14.66$ \\
 & {\modelstep} {\modelnumgroundinputunifs} {\modelnumgroundoutputunifs} {\modelnumvaroutputunifs}  & $19.44$ \\
 & {\modelstep}  & $43.04$ \\
\whline
\end{tabular}
  \end{center}
  \caption{Global comparison of the accuracy of cost models.}
\label{global-comparative}
\vspace{-0.5cm}
\end{table}

We have tested the implementation of the proposed cost models in order
to assess how well they predict the execution time of other programs (not
used in the calibration process) statically, without performing any
runtime profiling with them.  We have performed experiments with all
of the 63 possible cost models that result of the combination of one or
more of the components described in
Section~\ref{sec:proposed-cost-models}.  However, for space reasons
and for clarity, we only show the three most accurate cost models
(according to a global accuracy comparison that will be presented later) plus
the {\modelstep} model, which has special interest as we will also see
later. Experimental results are shown in Table~\ref{predict-mean-consider_builtins}.
{\bf Prog.} lists the program names. The analyzers integrated in the
\ciaopp\ system infer the exact cost function for all the programs in
that table under the $\indepmetric(\costmodel_i)$ metric, which means
that the upper and lower bound are the same, i.e.,  ${\tt
  Cost_{p}}^{l}(\indepmetric(\costmodel_i), \overline{n}) = {\tt
  Cost_{p}}^{u}(\indepmetric(\costmodel_i), \overline{n}) = {\tt
  Cost_p}(\indepmetric(\costmodel_i), \overline{n})$.  There are
several rows for each program in the table. The first three rows show
results corresponding to the prediction of execution times with the
three more accurate cost models. The fourth row shows the prediction
obtained by the cost model $\modelstep$ that only considers resolution
steps, i.e., it assumes that the execution time of a procedure call is
directly proportional to the number of resolution steps performed by
the call.  This means that for this simple cost model we are assuming
that $\headtime(\modelstep) = \constantstep$, since $I(\modelstep) =
1$, for a constant $\constantstep$, which represents the time taken by
a resolution step. Note that ${\tt
  Cost_{C}}(I(\modelstep),\overline{n})$ gives the number of
resolution steps performed by clause {\tt C}.  The last row per
benchmark program presents the observed execution times (i.e.,
measured execution times) and allows measuring the accuracy of the
different predictions. In this sense, values in the {\bf Model} column
are the names of the four cost models. The value {\bf observed}
identifies the row corresponding to the observed values.  The
following two columns show results corresponding to the
``\platformintel'' and ``\platformppc'' execution platforms. 

Column {\bf Estimate} shows execution times computed by
using the average value of the constant $\headconstantvector$ as estimated
in Table~\ref{average-k-only-unif}:

\[ 
 {\bf Estimate} = \headconstantvector \bullet
  \overline{{\tt Cost_p}}(\indepmetricvector, \overline{n})
\]

\noindent
Deviations respect to the observed values (in the {\bf observed} row) are
also shown between parenthesis in the column {\bf Estimate}.

The observed execution times have been measured by running the
programs with input data of a fixed size. {\numvaluesperexample} input
data sets of such fixed size have been generated randomly.
{\callspermeasure} runs of the program have been performed for each 
such input data set. The observed execution time for such input size has
been computed as the average of all runs.

Row ${\timeanalysis}$ shows the total (static) cost analysis
time (in seconds) needed to perform the execution time estimation (and
includes mode, type, and cost analysis).

Table~\ref{global-comparative} compares the overall accuracy of the
four cost models already shown in
Table~\ref{predict-mean-consider_builtins}, for the two considered
platforms.  The last column shows the global error and it is an
indicator of the amount of deviation of the execution times estimated
by each cost model with respect to the observed values. 
As global error we take the square mean of the errors in each example
being considered in Table~\ref{predict-mean-consider_builtins}.
By considering both platforms in combination we can conclude that the
more accurate cost model is the one consisting of steps,
{\modelnumgroundinputunifs}, {\modelnumgroundoutputunifs},
{\modelnumvarinputunifs}, and {\modelnumvaroutputunifs}.  This cost
model has an overall error of 14.66 \% in platform ``PPC'' and 31.06
\% in ``Intel''.  In ``Intel'' (obviously a more challenging platform)
the model consisting of steps, {\modelnumgroundinputunifs},
{\modelnumgroundoutputunifs}, and {\modelnumvaroutputunifs} appears to
be the best.
This coincides with our intuition that taking into account a
comparatively large number of lower-level operations should improve
accuracy. However, such components should contribute significantly
to the model in order to avoid noise introduction.
It is also interesting to see that including $\modelnumargs$ in the
cost model does not further improve accuracy, as expected, since
{\modelnumargs} is not independent from the four components
{\modelnumgroundinputunifs}, {\modelnumgroundoutputunifs},
{\modelnumvarinputunifs}, {\modelnumvaroutputunifs}. In fact,
including this component results in a less precise model in both
platforms, due to the noise introduced in the model.  Also, the cost
model \modelstep\ deserves special mention, since it is the simplest
one and, at least for the given examples, the error is smaller than we
expected and better than more complex cost models not shown in the
tables.

Overall we believe that the results are very encouraging in the sense
that our combined framework predicts with an acceptable
degree of accuracy the execution times of programs and paves the way
for even more accurate analyses by including additional parameters.

\section{Applications}
\label{sec:applications}

The experimental results presented in
Section~\ref{sec:asses:prediction}  show that the proposed
framework can be relevant in practice for estimating platform
dependent cost metrics such as execution time.
We believe that execution time estimates can be very useful in several
contexts. As already mentioned, in certain mobile/pervasive
computation scenarios different platforms come into play, with each
platform having different capabilities.  More concretely, the
execution time estimates could be useful for performing
resource/granularity control in parallel/distributed computing. This
belief is based on previous experimental results, where it appeared
from the sensitivity of the results observed in such experiments, that
while it is not essential to be absolutely precise in inferring the
best time estimates for a query, the number of reductions by itself was a
rough measure and the current time estimation approach could
presumably improve on previous results.

One of the good features of our approach is that we can translate
platform-independent cost functions (which are the result of the
analyzer) into platform-dependent cost functions (using the
relationship in expression~(\ref{eq:costrelation-vector})).
A possible application for taking advantage of this feature is mobile
code safety and in particular Proof-Carrying Code (PCC), a general
approach in which the code supplier augments the program with a
certificate (or proof). Consider a scenario where the producer sends a
certificate with a platform-independent cost function (i.e., where the
cost is expressed in a platform-independent metric) together with a
calibration program. The calibration program includes a fixed set of
calibration benchmarks. Then, the consumer runs (only once) the
calibration program and computes the values for the constants
appearing in the cost functions.
Using these constants, the consumer can obtain platform-dependent cost
functions~\cite{acc-res-ppdp05}.

\forlongvers{Details can be worked out to ensure that the consumer
  runs only once the calibration program.  Perhaps, there could be an
  initial communication between the producer and consumer so that the
  consumer runs only once the calibration program and ends up with the
  set of constants and the translator program. After that, the
  consumer only needs to send platform-independent cost functions.

The calibration program could have "state" so that if it is called a
second time then the calibration process is not run again but the
constants are returned. There must be a way to reset the calibrator.

Another scenario can be one where the producer and the consumer have a
previous communication, so that the consumer runs the calibration
program and sends the vector of constants to the producer. Then, the
producer sends platform-dependent cost functions to each particular
consumer. The producer has a table of consumers together with their
vectors of constants, so that the messages sent by the producer are
addressed to a particular consumer.

Another thing which should be taken into account is that if the
calibration program comes from the producer, then it is untrusted
code. Another certificate should be used for the calibration program.
This is needed in order to make sure that the platform-dependent
constants are correctly computed.  }

Another application of the proposed approach is
resource-oriented specialization. The proposed cost-models, which
include low-level factors for CLP programs, are more refined
cost-models than previously proposed ones and thus can be used to
better guide the specialization process. The inferred cost functions
can be used to develop automatic program transformation techniques
which take into account the size of the resulting program, its run
time and memory usage, and other low-level implementation factors. In
particular, they can be used for performing self-tuning
specialization in order to compare different specialized version
according to their costs~\cite{CraigLeuschel:PPDP05-short}.

\section{Conclusions}
\label{section:conclusions}

We have developed a framework which allows estimating 
execution times of procedures of a program in a given execution
platform. The method proposed combines compile-time (static) cost
analysis with a one-time profiling of the platform in order to
determine the values of certain constants. These constants calibrate a
cost model from which time cost functions 
for a given platform can be computed statically. The approach has been
implemented and integrated in the \ciaopp\ system.  To the best of our
knowledge, this is the first combined framework for estimating
statically  and accurately execution time bounds based on static
automatic inference of upper and lower bound complexity functions plus
experimental adjustment of constants.
We have performed an experimental assessment of this implementation
for a wide range of different candidate cost models and two execution
platforms.  The results achieved show that the combined framework
predicts the execution times of programs with a reasonable degree
of accuracy. We believe this is an encouraging result,
since
using a one-time profiling for estimating execution times of other,
unrelated programs is clearly a challenging goal.

Also, we argue that the work presented in this paper presents an
interesting trade-off between accuracy and simplicity of the approach.
At the same time, there is clearly room for improving precision by using more
refined cost models which take into account additional (lower level)
factors. Of course, these models would also be more difficult to handle 
since on one hand they would require computing more constants
and on the other hand they may require taking into account factors
which are not observable at source level. This is in any case the
subject of possibly interesting future work.

\section*{Acknowledgments}
This work was funded in part by the Information Society Technologies
program of the European Commission, Future and Emerging Technologies
under the IST-15905 {\em MOBIUS} project, by the Spanish
Ministry of Education under the TIN-2005-09207 {\em MERIT} project,
and the Madrid Regional Government under the \emph{PROMESAS} project.
Manuel Hermenegildo is also supported by the Prince of Asturias Chair
in Information Science and Technology at UNM.

\bibliographystyle{plain}

\begin{thebibliography}{10}

\bibitem{Brassel04LOPSTR}
B.~Brassel, M.~Hanus, F.~Huch, J.~Silva, and G.~Vidal.
\newblock Run-time profiling of functional logic programs.
\newblock In {\em Proceedings of the International Symposium on Logic-based
  Program Synthesis and Transformation (LOPSTR'04)}, pages 182--197. Springer
  LNCS 3573, 2005.

\bibitem{CraigLeuschel:PPDP05-short}
S.J. Craig and M.~Leuschel.
\newblock Self-tuning resource aware specialisation for {P}rolog.
\newblock In {\em Proc. of PPDP'05}, pages 23--34. ACM Press, 2005.

\bibitem{Debray83}
S.~K. Debray.
\newblock Profiling prolog programs.
\newblock {\em Software Practice and Experience}, 18(9):821--839, 1983.

\bibitem{granularity}
S.K. Debray, N.-W. Lin, and M.~Hermenegildo.
\newblock {T}ask {G}ranularity {A}nalysis in {L}ogic {P}rograms.
\newblock In {\em Proc. of the 1990 {ACM} Conf. on Programming Language Design
  and Implementation}, pages 174--188. {ACM} Press, June 1990.

\bibitem{caslog}
S.K. Debray and N.W. Lin.
\newblock Cost analysis of logic programs.
\newblock {\em {ACM} Transactions on Programming Languages and Systems},
  15(5):826--875, November 1993.

\bibitem{low-bounds-ilps97}
S.K. Debray, P.~L\'{o}pez-Garc\'{\i}a, M.~Hermenegildo, and N.-W. Lin.
\newblock {L}ower {B}ound {C}ost {E}stimation for {L}ogic {P}rograms.
\newblock In {\em 1997 International Logic Programming Symposium}, pages
  291--305. MIT Press, Cambridge, MA, October 1997.

\bibitem{Ducasse99}
Mireille Ducass{\'e}.
\newblock Opium: An extendable trace analyzer for prolog.
\newblock {\em J. Log. Program.}, 39(1-3):177--223, 1999.

\bibitem{acc-res-ppdp05}
M.~Hermenegildo, E.~Albert, P.~L\'{o}pez-Garc\'{\i}a, and G.~Puebla.
\newblock {A}bstraction {C}arrying {C}ode and {R}esource-{A}wareness.
\newblock In {\em Proc. of PPDP'05}. ACM Press, July 2005.

\bibitem{ciaopp-sas03-journal-scp-short}
M.~Hermenegildo, G.~Puebla, F.~Bueno, and P.~L\'{o}pez-Garc\'{\i}a.
\newblock {I}ntegrated {P}rogram {D}ebugging, {V}erification, and
  {O}ptimization {U}sing {A}bstract {I}nterpretation (and {T}he {C}iao {S}ystem
  {P}reprocessor).
\newblock {\em Science of Computer Programming}, 58(1--2):115--140, October
  2005.

\bibitem{pedro-phd}
P.~L\'{o}pez-Garc\'{\i}a.
\newblock {\em Non-failure Analysis and Granularity Control in Parallel
  Execution of Logic Programs}.
\newblock PhD thesis, Universidad Polit\'{e}cnica de Madrid (UPM), Facultad
  Informatica UPM, 28660-Boadilla del Monte, Madrid-Spain, June 2000.

\bibitem{granularity-jsc-short}
P.~L\'{o}pez-Garc\'{\i}a, M.~Hermenegildo, and S.K. Debray.
\newblock {A} {M}ethodology for {G}ranularity {B}ased {C}ontrol of
  {P}arallelism in {L}ogic {P}rograms.
\newblock {\em J. of Symbolic Computation, Special Issue on Parallel Symbolic
  Computation}, 22:715--734, 1996.

\bibitem{morgan98}
S.~A.~Jarvis R.~G.~Morgan.
\newblock Profiling large-scale lazy functional programs.
\newblock {\em Journal of Functional Programing}, 8(3):201--237, May 1998.

\bibitem{sansom97formally}
Patrick~M. Sansom and Simon L.~Peyton Jones.
\newblock Formally based profiling for higher-order functional languages.
\newblock {\em ACM Transactions on Programming Languages and Systems},
  19(2):334--385, March 1997.

\bibitem{mendenhall95}
D.~Wackerly, W.~Mendenhall, and R.~Scheaffer.
\newblock {\em Mathematical Statistics With Applications 5th Edition}.
\newblock P W S Publishers, 1995.

\end{thebibliography}

\newpage

\end{document}